\documentclass[twocolumn,showpacs,preprintnumbers,amsmath,amssymb,superscriptaddress]{revtex4}

\usepackage[dvipdfm]{graphicx}
\begin{document}
\title{Electronic structure of Ga$_{1-x}$Cr$_{x}$N and Si-doping effects studied by photoemission and X-ray absorption spectroscopy}

\author{G.~S.~Song}
\affiliation{Department of Physics and Department of Complexity Science and Engineering, University of Tokyo, 
7-3-1 Hongo, Bunkyo-ku, Tokyo, 113-0033, Japan}
\author{M.~Kobayashi}
\affiliation{Department of Physics and Department of Complexity Science and Engineering, University of Tokyo, 
7-3-1 Hongo, Bunkyo-ku, Tokyo, 113-0033, Japan}
\author{J.~I.~Hwang}
\affiliation{Department of Physics and Department of Complexity Science and Engineering, University of Tokyo, 
7-3-1 Hongo, Bunkyo-ku, Tokyo, 113-0033, Japan}
\author{T.~Kataoka}
\affiliation{Department of Physics and Department of Complexity Science and Engineering, University of Tokyo, 
7-3-1 Hongo, Bunkyo-ku, Tokyo, 113-0033, Japan}
\author{M.~Takizawa}
\affiliation{Department of Physics and Department of Complexity Science and Engineering, University of Tokyo, 
7-3-1 Hongo, Bunkyo-ku, Tokyo, 113-0033, Japan}
\author{A.~Fujimori}
\affiliation{Department of Physics and Department of Complexity Science and Engineering, University of Tokyo, 
7-3-1 Hongo, Bunkyo-ku, Tokyo, 113-0033, Japan}
\affiliation{Synchrotron Radiation Research Unit, Japan Atomic Energy
Agency, Sayo-gun, Hyogo 679-5148, Japan}

\author{T.~Ohkouchi}
\affiliation{Synchrotron Radiation Research Unit, Japan Atomic Energy
Agency, Sayo-gun, Hyogo 679-5148, Japan}
\author{Y.~Takeda}
\affiliation{Synchrotron Radiation Research Unit, Japan Atomic Energy
Agency, Sayo-gun, Hyogo 679-5148, Japan}
\author{T.~Okane}
\affiliation{Synchrotron Radiation Research Unit, Japan Atomic Energy
Agency, Sayo-gun, Hyogo 679-5148, Japan}
\author{Y.~Saitoh}
\affiliation{Synchrotron Radiation Research Unit, Japan Atomic Energy
Agency, Sayo-gun, Hyogo 679-5148, Japan}
\author{H.~Yamagami}
\affiliation{Synchrotron Radiation Research Unit, Japan Atomic Energy
Agency, Sayo-gun, Hyogo 679-5148, Japan}
\affiliation{Department of Physics, Faculty of Science,
Kyoto Sangyo University, Kyoto 603-8555, Japan}

\author{F.-H.~Chang}
\affiliation{National Synchrotron Radiation Research Center, Hsinchu 30076,
Taiwan}
\author{L.~Lee}
\affiliation{National Synchrotron Radiation Research Center, Hsinchu 30076,
Taiwan}
\author{H.-J.~Lin}
\affiliation{National Synchrotron Radiation Research Center, Hsinchu 30076,
Taiwan}
\author{D.~J.~Huang}
\affiliation{National Synchrotron Radiation Research Center, Hsinchu 30076,
Taiwan}
\author{C.~T.~Chen}
\affiliation{National Synchrotron Radiation Research Center, Hsinchu 30076,
Taiwan}

\author{S.~Kimura}
\affiliation{The Institute of Scientific and Industrial Research, Osaka University, 8-1 Mihogaoka, Ibaraki, Osaka 567-0047, Japan}
\author{M.~Funakoshi}
\affiliation{The Institute of Scientific and Industrial Research, Osaka University, 8-1 Mihogaoka, Ibaraki, Osaka 567-0047, Japan}
\author{S.~Hasegawa}
\affiliation{The Institute of Scientific and Industrial Research, Osaka University, 8-1 Mihogaoka, Ibaraki, Osaka 567-0047, Japan}
\author{H.~Asahi}
\affiliation{The Institute of Scientific and Industrial Research, Osaka University, 8-1 Mihogaoka, Ibaraki, Osaka 567-0047, Japan}

\date{\today}

\begin{abstract}
The electronic structure of the magnetic semiconductor Ga$_{1-x}$Cr$_{x}$N and the effect of Si doping on it have been investigated by photoemission and soft x-ray absorption spectroscopy. We have confirmed that Cr in GaN is predominantly trivalent substituting for Ga, and that Cr 3$d$ states appear within the band gap of GaN just above the N 2$p$-derived valence-band maximum. As a result of Si doping, downward shifts of the core levels (except for Cr 2$p$) and the formation of new states near the Fermi level were observed, which we attribute to the upward chemical potential shift and the formation of a small amount of Cr$^{2+}$ species caused by the electron doping. Possibility of Cr-rich cluster growth by Si doping are discussed based on the spectroscopic and magnetization data. 
\end{abstract}


\maketitle

 Diluted magnetic semiconductors (DMSs) with Curie temperatures ($T\mathrm{_C}$'s) significantly above room temperature are needed for the development of spintronics devices for practical applications. 
 Ga$_{1-x}$Cr$_{x}$N has been predicted theoretically to be ferromagnetic via double exchange (DE) mechanism \cite{KSato_GaN1} and indeed ferromagnetism (FM) above room temperature has been observed \cite{HashimotoJCG03, JJKim_pssc}. 
 X-ray absorption and photoemission studies have revealed strong hybridization between the Cr 3$d$ states and the valence $p$ band \cite{Hashimoto_KXAS}, and the formation of Cr 3$d$-derived states within the band gap of GaN \cite{JJKim_3p3d, Hwang_GCN}, which suggested a high potential as an intrinsic DMS. However, high resistivity \cite{JJKim_JVSTB} and the absence of density of states at Fermi level ($E\mathrm{_F}$) \cite{JJKim_3p3d, Hwang_GCN} have thrown doubt on the DE mechanism of the FM in Ga$_{1-x}$Cr$_{x}$N. 
 On the other hand, the substitutional transition-metal (TM)-rich clusters of nanometer sizes embedded in the host semiconductor have been observed in Zn$_{1-x}$Cr$_{x}$Te \cite{Kuroda} and suggested as another origin of ferromagnetism in DMSs \cite{KSato_JJAP05}. Such TM-rich clusters have also been observed in other III-V-based wide-gap DMSs, Ga$_{1-x}$Mn$_{x}$N \cite{Dhar_APL031}, In$_{1-x}$Cr$_{x}$N \cite{Lin_JMMM051}, and Ga$_{1-x}$Fe$_{x}$N \cite{Bonanni}.

In order to clarify the mechanism of the FM in Ga$_{1-x}$Cr$_{x}$N, whether the DE mechanism or the Cr-rich cluster formation, the investigation of the carrier-concentration dependence of the electronic structure together with the magnetic properties is expected to yield essential information as in the case of N- and I-doped Zn$_{1-x}$Cr$_{x}$Te \cite{Ozaki_Ndope, Ozaki_Idope}. 
 Recently, Funakoshi $et$ $al$.~\cite{Funakoshi} reported a successful control of carrier concentration in Ga$_{1-x}$Cr$_{x}$N by doping Si, which is expected to replace Ga and hence acts as a donor. 
In this work, we report on photoemission (PES) and x-ray absorption spectroscopy (XAS) studies of Si-doping effect on the electronic structure of Ga$_{1-x}$Cr$_{x}$N samples whose magnetic properties have been characterized. The results indicate that Si doping causes an upward chemical potential shift and the formation of new states near $E\mathrm{_F}$ (presumably derived from Cr$^{2+}$). 
We suggest that Cr-rich cluster formation may be enhanced by Si doping, and may stabilize FM or superparamagnetism (SPM) in Ga$_{1-x}$Cr$_{x}$N. Possible origins of the observed FM or SPM shall be discussed below.

 Ga$_{1-x}$Cr$_{x}$N thin films were grown by radio-frequency (rf) plasma-assisted molecular beam epitaxy on 2 $\mu$m GaN-template substrates which had been grown by metal-organic chemical vapor deposition. Elemental Ga, Cr, Si, and rf plasma-enhanced N$_{2}$ were used as sources. After thermally cleaning the substrate surface at 700 $^{\circ}$C for 15 min, a high-temperature GaN buffer layer was grown at 700 $^{\circ}$C. Finally, a Ga$_{0.98}$Cr$_{0.02}$N (GaCrN) or Si-doped Ga$_{0.98}$Cr$_{0.02}$N (GaCrN:Si) layer was grown at a relatively low temperature of 700 $^{\circ}$C in order to increase the Cr solubility. The Ga flux, N$_{2}$ flow rate, Cr cell temperature, and Si cell temperature were set at 1.6$\times$10$^{-7}$ Torr, 1.5 SCCM, 980 $^{\circ}$C, and 1100 $^{\circ}$C, respectively. The nominal carrier concentrations were $1\times10^{18}$ and $2\times10^{19}$ /cm$^{3}$ for GaCrN and GaCrN:Si, respectively, and the Si concentration should be somewhat larger than them.
The thickness of the Cr-doped layer was about 19 nm. For comparison, Si-doped GaN without Cr doping (GaN:Si) was also grown by the same procedure. To avoid the oxidation of sample surfaces and to perform the PES and XAS measurements without surface cleaning, 2 nm-thick capping layers (GaN for GaCrN and GaN:Si for GaCrN:Si, respectively) were deposited on the sample surfaces. Magnetization was measured using a SQUID magnetometer (MPMS $XL$, Quantum Design, Co., Ltd.). 

X-ray photoemission spectroscopy (XPS) measurements were performed using a Mg-$K\alpha$ source and a Gammadata Scienta SES-100 hemispherical analyzer. 
Resonant photoemission spectroscopy (RPES) measurements were performed at BL-23SU of SPring-8 using synchrotron radiation. Photoelectrons were collected using a Gammadata Scienta SES-2002 hemispherical analyzer. The photoemission spectra were referenced to the Au 4$f$ peak and the $E\mathrm{_F}$ of gold which was in electrical contact with the samples for XPS and RPES, respectively.
 XAS measurements were performed at the Dragon Beamline BL-11A of National Synchrotron Radiation Research Center in the total-electron-yield mode.
All the spectra were taken at room temperature. The total resolution of the spectrometer including the temperature broadening was $\sim$800 meV and $\sim$200 meV for XPS and RPES, respectively. The monochromator resolution for XAS was $E$/$\Delta$$E$ $>$ 10,000. The base pressure of spectrometer was below 3$\times$10$^{-8}$ Pa. 

\begin{figure}[t]
\begin{center}
\includegraphics[width=\linewidth]{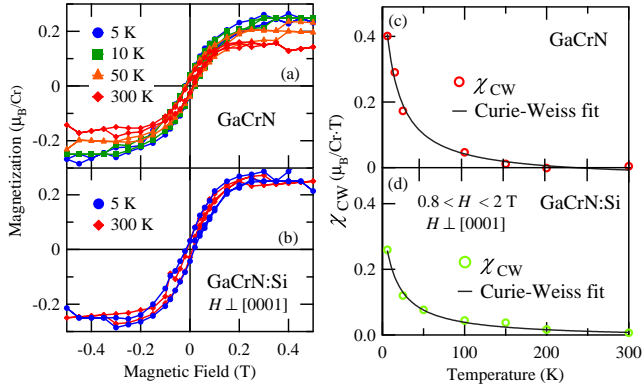}
\caption{(Color Online) Magnetization of GaCrN and GaCrN:Si. (a), (b) $M$-$H$ curves at various temperatures. (c), (d) Curie-Weiss (CW) component of the high-field magnetic susceptibility as a function of temperature.}
\label{MPMS_GCN}
\end{center}
\end{figure}

The magnetization data of the two Cr-doped samples are shown in Fig.~\ref{MPMS_GCN}. Panels (a) and (b) shows $M$-$H$ curves at various temperatures. Here, the linear components have been subtracted. Such a $M$-$H$ curve is known as an characteristic feature of SPM. One can see that the $T\mathrm{_C}$ exceeds 400 K for both samples. 
The saturation magnetization of the GaCrN slightly decreases with increasing temperature, while that of the Si-doped one hardly decreases with temperature, as shown in panels (a) and (b), respectively. Since, in general, saturation magnetization is constant on temperature far below $T\mathrm{_C}$, the observation suggests that Si doping in Ga$_{1-x}$Cr$_{x}$N causes an increase of $T\mathrm{_C}$. 
Figure \ref{MPMS_GCN}(c) and (d) shows the temperature dependence of high-field (0.8~$<$~$H$~$<$~2 T) magnetic susceptibility and Curie-Weiss (CW) fit. The fitting was made assuming a CW term plus a temperature independent constant ${\partial M}/{\partial H}~=~NC/(T-\Theta) + \chi_0$. Here, $N$ is number of the magnetic ions, $g$ is the $g$ factor, $C~=~(g\mu_B)^2S(S+1)/3k_B$ is the Curie constant (assuming $S$~=~3/2 and $g$~=~2), and $\theta$ is the Weiss temperature. After the fitting, the $\chi_0$ term has been removed. Fractions of the Cr atoms showing a CW-paramagnetic behavior are $\sim$26\% for GaCrN and $\sim$16\% for GaCrN:Si. 
The decrease of the CW component with Si doping indicates that the number of isolated (paramagnetic) Cr atoms decreases, probably because they were absorbed by the Cr-rich clusters. 

\begin{figure}[bp]
\begin{center}
\includegraphics[width=\linewidth]{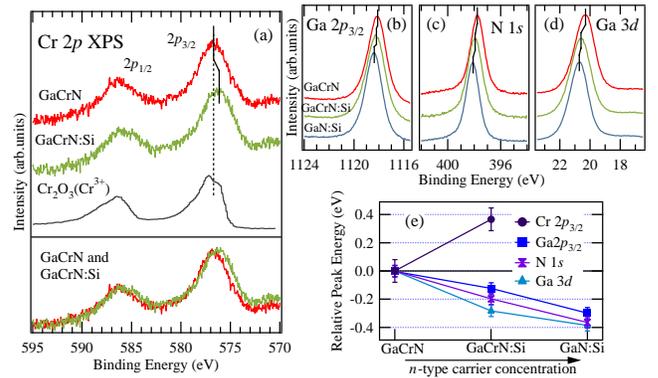}
\caption{(Color Online) Core-level XPS spectra of GaCrN, GaCrN:Si and Si-doped GaN (GaN:Si). (a) Cr 2$p$, (b) Ga 3$d$, (c) N 1$s$, and (d) Ga 2$p_{3/2}$ core levels. Electron mean free paths in GaP for the Cr 2$p$, Ga 3$d$, N 1$s$, and Ga 2$p_{3/2}$ core levels are $\sim$2.0, $\sim$3.1, $\sim$2.3, and $\sim$0.7 nm, respectively \cite{Mfp}. Vertical lines represent the peak position of each core level. In (a), the Cr 2$p$ core-level XPS spectrum of Cr$_{2}$O$_{3}$ \cite{Cr2O3XPS} is also shown for comparison. The spectra of GaCrN and GaCrN:Si are overlaid at the bottom panel. (e) Shifts of the core-level peaks of GaCrN:Si and GaN:Si relative to those of GaCrN. The $n$-type carrier concentration increases in order of GaCrN, GaCrN:Si, GaN:Si.}
\label{XPS_GCN}
\end{center}
\end{figure}

 Figure \ref{XPS_GCN}(a)-(d) shows the core-level XPS spectra of GaCrN, GaCrN:Si, and GaN:Si. In panel (a), the two peaks in each spectrum are due to the Cr 2$p_{3/2}$-2$p_{1/2}$ spin-orbit doublet. The Cr 2$p$ spectrum of Cr$_{2}$O$_{3}$ (Cr$^{3+}$) \cite{Cr2O3XPS} is also shown for comparison. The peak position for GaCrN is close to that for Cr$_{2}$O$_{3}$, indicating that the valence of Cr in GaCrN is close to 3+, that is, the Cr atom is doped into GaN as a neutral impurity if Cr substitutes for Ga, consistent with the earlier reports \cite{JJKim_JVSTB, Hwang_GCN, Hashimoto_JAP06}. The core-level peaks shown in panels (b)-(d) are shifted downwards in GaCrN:Si and GaN:Si compared with GaCrN, indicating that the Si doping causes an upward shift of the chemical potential $\mu$. Furthermore, one notices that the core-level peaks of GaCrN:Si are shifted upwards compared with GaN:Si, suggesting that Cr acts as an acceptor and compensates the electron carriers, consistent with the previous reports that Cr doping make semiconducting GaN samples insulating \cite{JJKim_JVSTB, Shanthi_JJAP}. The amount of the peak shift of each core level relative to that in GaCrN is plotted in panel (e). The effect of electron trapping by Cr appears in the Cr 2$p$ spectrum of GaCrN:Si as shown at the bottom of panel (a): one can see the formation of a small hump for the Si-doped sample and the opposite peak shift to the other core levels. In general, the binding energy ($E_B$) of a core-level spectrum decreases with decreasing the valence of the element due to the increased screening effect by valence electrons. We therefore consider that electrons supplied by the Si doping are trapped at Cr site and convert those atoms from Cr$^{3+}$ to Cr$^{2+}$.

\begin{figure}[bp]
\begin{center}
\includegraphics[width=\linewidth]{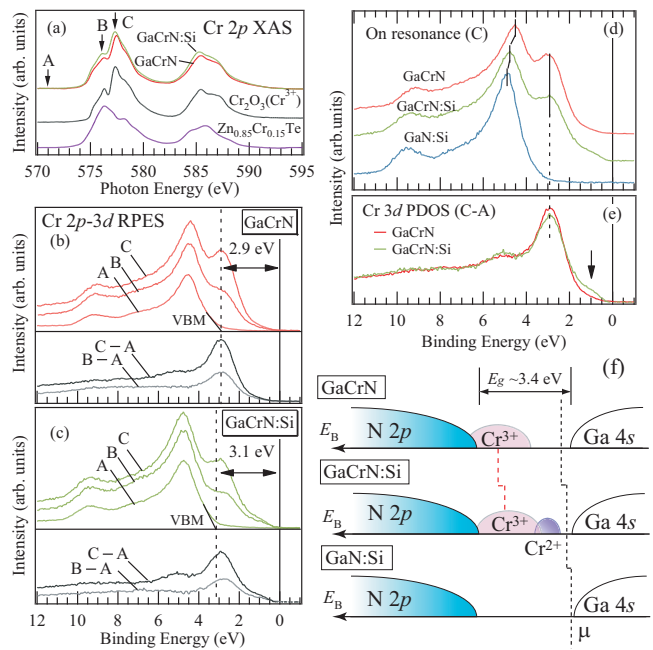}
\caption{(Color Online) Cr 2$p$ XAS spectra (a) and Cr 2$p$-3$d$ RPES spectra of GaCrN (b) and GaCrN:Si (c) taken with photon energies A, B and C ($h\nu$ : A $\sim$571.5 eV, B $\sim$576 eV, C $\sim$577.5 eV) as shown in panel (a). The difference spectra B-A and C-A are shown at the bottom of panel (b) and (c). The VBMs estimated from the off-resonance spectra are indicated by vertical dashed lines. (d) Comparison of the on-resonance spectra of the different samples taken with photon energy C. The spectra have been normalized at the peak intensity of the N 2$p$-derived valence band at $E_B$ $\sim$4.5-5 eV. (e) Si-doping dependence of the Cr 3$d$ PDOS. Arrow indicates a new Cr 3$d$ PDOS feature induced by the Si doping. (f) Schematic diagram of the electronic structures of GaCrN, GaCrN:Si, and GaN:Si. The shifts of chemical potential $\mu$, the upward shift of Cr 3$d$ PDOS peak, and the formation of the new states of Cr$^{2+}$ origin are shown. Here, the band gap $\sim$3.4 eV of hexagonal GaN is also indicated.}
\label{RPES_GCN}
\end{center}
\end{figure}

 The Cr 2$p$ XAS spectra of GaCrN and GaCrN:Si are shown in Fig.~\ref{RPES_GCN}(a). For comparison, those of Cr$_{2}$O$_{3}$ and Zn$_{1-x}$Cr$_{x}$Te \cite{ZnCrTe_Kobayashi} are also shown. 
The peak positions and the line shape of the spectrum of GaCrN agree with those of Cr$_{2}$O$_{3}$ (Cr$^{3+}$), and thus one can conclude that the valence of Cr in GaCrN and GaCrN:Si is predominantly 3+. Comparing the spectra of GaCrN and GaCrN:Si, one can see a slight difference at $\sim$577 eV and 585 eV. 
The small difference implies that only a small amount of Cr$^{2+}$ species exist. The difference corresponds well to the overall spectral weight distribution of Zn$_{1-x}$Cr$_{x}$Te (Cr$^{2+}$). 
 Figure \ref{RPES_GCN}(b) and (c) shows the valence-band RPES spectra of GaCrN and GaCrN:Si recorded in the Cr 2$p$-3$d$ excitation region. 
The difference spectra obtained by subtracting the off-resonance spectrum (A) from the on-resonance spectra (B or C) shown at the bottom of panels (b) and (c) represent the Cr 3$d$ partial density of states (PDOS). One can clearly see that Cr 3$d$-derived emission is seen mostly within the band gap of GaN as reported in Refs.~\cite{JJKim_3p3d, Hwang_GCN}. Also the existence of Cr-related states at $\sim$340 meV above the valence-band maximum (VBM) was reported in Ref. \cite{Shanthi_JJAP} from a photoluminescence study. The peak position of the Cr 3$d$ PDOS of GaCrN relative to the VBM well agrees with this value. 
Here, we have estimated the position of the VBM by the intersection between the zero level and a linear extrapolation of the low $E_B$ edge of the valence-band spectrum \cite{VBM}. The distance between the $E\mathrm{_F}$ and VBM was estimated to be $\sim$2.9 eV for GaCrN and $\sim$3.1 eV for GaCrN:Si as shown panels (b) and (c).

Figure \ref{RPES_GCN}(d) and (e) shows comparison of the on-resonance spectra (C) of the three samples and comparison of the Cr 3$d$ PDOS (C-A) of GaCrN and GaCrN:Si.
In panel (e), one clearly finds that new states are formed around $E_B$ $\sim$1 eV by the Si doping, as indicated by a vertical arrow. Considering the results of XPS and XAS, we attribute this new feature to a Cr$^{2+}$ origin. The new states are created as deep as $E\mathrm{_B}\sim$1 eV in spite of the small chemical potential shift of $\sim$200 meV, quite different from the conventional rigid-band behavior. While the entire valence band is shifted towards higher $E_B$ by Si doping, the Cr 3$d$ PDOS stays at the same $E_B$, as indicated by a vertical dashed line in panel (d). The formation of the new states and the absence of shifts in the Cr 3$d$ PDOS may be explained by the strong Coulomb repulsion between the Cr 3$d$ electrons \cite{Morikawa}. We note that no Cr 3$d$ PDOS was observed at $E\mathrm{_F}$. 
 In Fig.~\ref{RPES_GCN}(f), we show a schematic diagram of the electronic structure of GaCrN, GaCrN:Si, and GaN:Si where the chemical potential shift, the upward shift of Cr 3$d$ PDOS, and the formation of the new states are presented. 

 In the preceding paragraphs, we have considered changes in the electronic structure by Si doping as a result of Cr$^{2+}$ formation. 
In order to discuss the relationship between the observed changes in the electronic structure and the magnetism of Ga$_{1-x}$Cr$_{x}$N caused by Si doping, we consider the formation of Cr-rich clusters as follows. 
If Cr ions are randomly distributed in GaN, magnetic interaction between Cr ions would be weak because the interaction would be short-ranged \cite{KSato_range}. 
Considering the results of the magnetization measurements, which imply the existence of Cr-rich clusters and its growth by Si doping, and the results of the PES and XAS measurements, which suggest the increase of Cr$^{2+}$ states with Si doping and the absence of Cr 3$d$ PDOS at $E\mathrm{_F}$, one can imagine that the coexistence of Cr$^{3+}$ and Cr$^{2+}$ ions within the Cr-rich cluster may induce short-range DE within the Cr-rich clusters and stabilize ferromagnetism. 
However, if there is ferromagnetic spin alignment within the Cr-rich clusters via DE, the saturation magnetization ($M\mathrm{_S}$) would increase with increasing cluster size significantly. In the present study, we could not observe such an increase of $M\mathrm{_S}$ and $M\mathrm{_S}$ remained small ($\sim$0.25 $\mu\mathrm{_B}$/Cr) as shown in Fig.~\ref{MPMS_GCN}(a) and (b). 
One possible reason for the absence of short-range DE is the insufficient amount of Cr$^{2+}$. Actually, the concentration of Si and Cr is $\sim10^{19}$cm$^{-3}$ and $\sim10^{21}$cm$^{-3}$, respectively, and therefore there are only few \% of Cr$^{3+}$ is converted to Cr$^{2+}$. 

Another scenario to explain the possible increase of $T\mathrm{_C}$ by Si doping is the growth of antiferromagnetic (AFM) Cr-rich clusters. 
Since the spin moment of the Cr$^{3+}$ ion is 3 $\mu\mathrm{_B}$, $M\mathrm{_S}$ $\sim$0.25 $\mu\mathrm{_B}$/Cr means that the amount of ferromagnetic Cr atoms in GaCrN and GaCrN:Si is only $\sim$8\%. The amount of paramagnetic Cr atoms estimated from the CW fit is $\sim$26\% and $\sim$16\% for GaCrN and GaCrN:Si, respectively. This indicates that $\sim$66\% and $\sim$76\% for GaCrN and GaCrN:Si the Cr atoms could not be detected by the present magnetization measurements and are probably antiferromagnetically coupled. Recently, Cui $et$ $al$.~\cite{Cui} have calculated the magnetic moment and the stability of substitutional Cr-rich clusters with several sizes in Ga$_{1-x}$Cr$_{x}$N based on first-principles density functional theory, and have shown that large Cr-rich clusters are energetically favored and Cr atoms in these large Cr-rich clusters are coupled antiferromagnetically. The calculated moments of the Cr-rich clusters with various sizes are 0.06-1.47 $\mu\mathrm{_B}$/Cr, corresponding well to the observed $M\mathrm{_S}$ $\sim$0.25 $\mu\mathrm{_B}$/Cr. Kim $et$ $al$.~\cite{JJKim_JVSTB} have reported that the $M\mathrm{_S}$ of Ga$_{1-x}$Cr$_{x}$N decreases with increasing Cr concentration, which can be interpreted as due to the size increase of the antiferromagnetic Cr-rich clusters. 
As for Zn$_{1-x}$Co$_{x}$O, for which even smaller $M\mathrm{_S}$ has been reported, Dietl $et$ $al$.~\cite{Dietl_ZnCoO} attributed the origin of the small $M\mathrm{_S}$ to the uncompensated moments of the Co-rich AFM nanoclusters. 
These theoretical and experimental reports lead us to speculate that the uncompensated moments from AFM Cr-rich clusters and the increase of the cluster size may be the origin of the observed magnetism and its Si-doping dependence in Ga$_{1-x}$Cr$_{x}$N. Also, the magnetic anisotrpy (factor of 2-3 in the saturation magnetization) observed for the present GaCrN and GaCrN:Si samples may be responsible for the observed apparent ferromagnetic benhavior.

We note that the enhancement of FM by donor atom doping is seen in I-doped Zn$_{1-x}$Cr$_{x}$Te \cite{Kuroda}. 
In the case of Zn$_{1-x}$Cr$_{x}$Te, in undoped samples, because ZnTe is known as a native $p$-type semiconductor, I-doping makes the Cr ion becomes neutral 2+. Kuroda $et$ $al$.~\cite{Kuroda} attributed the origin of the enhancement of FM to the Cr-rich cluster formation in the I-doped samples caused to attractive forces between neutral Cr impurities (Cr$^{2+}$) as opposed to the repulsive forces between charged Cr impurities (Cr$^{3+}$).  
In the present case, the Si doping makes a fraction of the Cr ions negatively charged, and therefore the charge neutrality cannot be a driving force of the Cr-rich cluster formation in the Si-doped GaN. 
Microscopy observation of Ga$_{1-x}$Cr$_{x}$N:Si are therefore highly desired in future studies.

 In summary, we have investigated the Si-doping dependence of the electronic structure of Ga$_{1-x}$Cr$_{x}$N and its relationship to magnetism using XPS, XAS, and RPES. The upward chemical potential shift induced by Si doping into Ga$_{1-x}$Cr$_{x}$N and downward one by Cr doping into GaN:Si were observed. The new states of Cr$^{2+}$ character were found to form near $E\mathrm{_F}$ in Cr 3$d$ PDOS spectra by the Si doping. The growth of AFM Cr-rich clusters is proposed to explain the observed change of magnetism by the Si doping.

This work was supported by a Grant-in-Aid for Scientific Research in Priority Area gCreation and Control of Spin Currenth(19048012) from MEXT, Japan. The experiment at SPring-8 was approved by the Japan Synchrotron Radiation Research Institute (JASRI) Proposal Review Committee (Proposal No. 2007A3832). We also thank the Material Design and Characterization Laboratory, Institute for Solid State Physics, University of Tokyo, for the use of the SQUID magnetometer.

\end{document}